\documentclass{PoS}

\usepackage{amsfonts,amssymb,amsthm,amsmath} 
\usepackage{graphicx}
\usepackage[utf8x]{inputenc}

\title{Renormalisation of the scalar energy-momentum tensor with the Wilson flow}

\ShortTitle{Renormalised scalar EMT with Wilson flow}

\newcommand{\edimburgo}{Higgs Centre for Theoretical Physics, School of Physics and Astronomy,
  University of Edinburgh, Edinburgh EH9 3FD, UK}
\newcommand{\plymouth}{Centre for Mathematical Sciences, Plymouth University, Plymouth PL4 8AA, UK}

\author{Francesco Capponi, Antonio Rago
	 \\
	 \plymouth\\
	 E-mail: \email{francesco.capponi@plymouth.ac.uk, antonio.rago@plymouth.ac.uk}}
	 
\author{Luigi Del Debbio, \speaker{Susanne Ehret}, Roberto Pellegrini, Antonin Portelli
	 \\
	 \edimburgo\\
	 E-mail: \email{luigi.del.debbio@ed.ac.uk, susanne.ehret@ed.ac.uk, r.pellegrini@ed.ac.uk, antonin.portelli@ed.ac.uk}}

\abstract{The non-perturbative computation of the energy-momentum tensor can be used to study the scaling behaviour of strongly coupled quantum field theories. The Wilson flow is an essential tool to find a meaningful formulation of the energy-momentum tensor on the lattice. We extend recent studies of the renormalisation of the energy-momentum tensor in four-dimensional gauge theory to the case of a three-dimensional scalar theory to investigate its intrinsic structure and numerical feasibility on a more basic level. In this paper, we discuss translation Ward identities, introduce the Wilson flow for scalar theory, and present our results for the renormalisation constants of the scalar energy-momentum tensor.}

\FullConference{34th annual International Symposium on Lattice Field Theory\\
		 24-30 July 2016\\
		 University of Southampton, UK}

\begin{document}

\newcommand{\tr}{\mbox{tr}} 
\newcommand{\e}{\mbox{e}}

\section{Introduction}

Our motivation to study the energy-momentum tensor (EMT) is the relation of its trace to the $\beta$-function via
\begin{equation}
 \langle \int d^Dx\, T_{\mu\mu}\, \phi(x_1) ... \phi(x_n) \rangle = -\left(\sum_k \beta_k \frac{\partial}{\partial g_k} + n(\gamma_{\phi} + d_{\phi}) \right) \langle \phi(x_1) ... \phi(x_n) \rangle,
 \label{eq:betaEMT}
\end{equation}
where $\beta_k$ are the beta functions, $g_k$ are the couplings of the theory, $\gamma_{\phi}=-\frac{1}{2Z_{\phi}}{\scriptstyle \mu}\frac{\text{d}}{\text{d}\mu} {\scriptstyle Z_{\phi}}$, $Z_{\phi}$ is the field renormalisation factor, and $d_{\phi}$ is the classical dimension of field $\phi$. Eq. (\ref{eq:betaEMT}) can be found using the Callan-Symanzik equation and the dilatation Ward identity. It enables us to investigate the scaling behaviour of the field theory under consideration. Scalar $\phi^4$-theory in three dimensions is particularly interesting as it exhibits an infrared fixed point in three dimensions for $m_0^2<0$, and can thus serve as a toy model for theories with such a fixed point.

The numerical renormalisation of the EMT was studied before, e.g. in \cite{Suzuki:2013gza,Giusti:2015daa}. The proposal to apply the Wilson flow to the computation of the EMT in Yang-Mills theory was made in \cite{DelDebbio:2013zaa}. We extend their investigation to the case of scalar $\phi^4$-theory to gain deeper insight into the theoretical background and to test the numerical feasibility of the method, see also \cite{Capponi:2015ucc}.

\begin{figure}[h]
\centering
  \includegraphics[width=6cm,trim={6cm 10.5cm 5.5cm 11.5cm},clip]{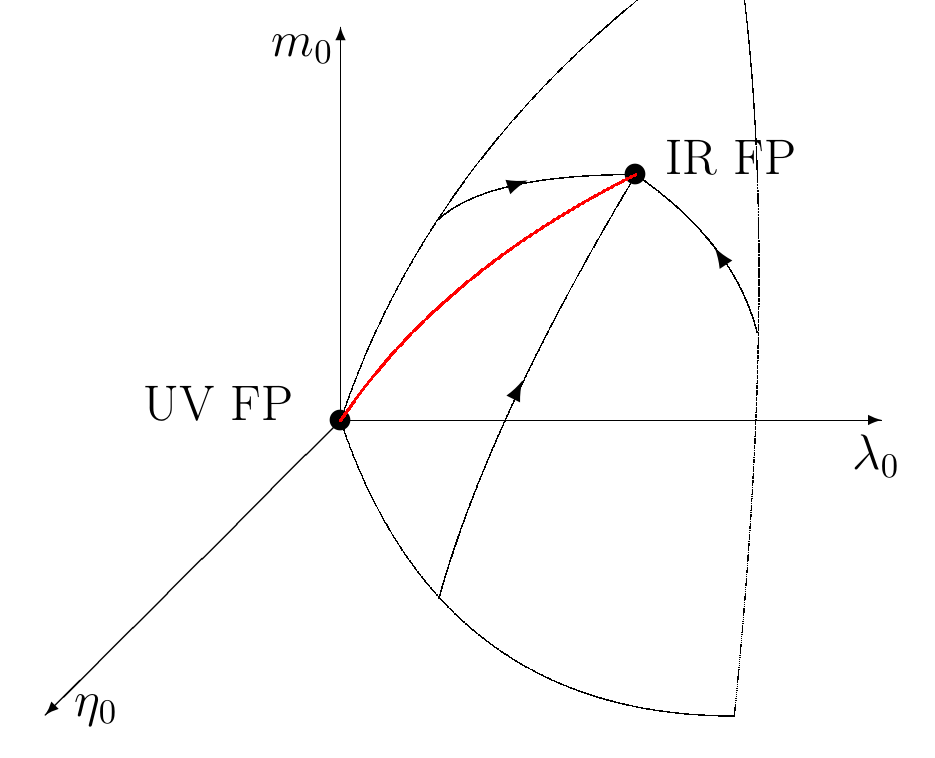}
  \caption{Sketch of the infinite coupling space of $\phi^4$-theory in three dimensions with $m_0^2<0$ showing the critical surface as well as the ultraviolet (UV) and infrared (IR) fixed points (FP).}
\label{fig:phasediag}
\end{figure}

The infinite coupling space of $\phi^4$-theory in three dimensions is sketched in fig. \ref{fig:phasediag}. There are only two relevant (bare) couplings, $m_0$ and $\lambda_0$. The direction $\eta_0$ represents all other couplings. The ultraviolet and infrared fixed points lie on the critical surface where the physical mass is zero. The critical surface separates the symmetric from the broken phase. The direction of the renormalisation group flow is indicated as well.

\section{Energy-momentum tensor in scalar field theory}

We consider scalar field theory in three dimensions with Euclidean action
\begin{equation}
 S = \int d^3x\; \left( \frac12(\partial_{\mu}\phi)^2 + \frac12m_0^2\phi^2 + \frac{\lambda_0}{4!}\phi^4 \right).
\end{equation}
The EMT is the collection of Noether currents associated to translations and can be found from the variation of the action,
\begin{equation}
 T_{\mu\rho}(x) = \partial_{\mu}\phi \partial_{\rho}\phi - \delta_{\mu\rho}\left( \frac12 \sum_\sigma\partial_{\sigma}\phi \partial_{\sigma}\phi + \frac12 m_0^2\phi^2 + \frac{\lambda_0}{4!} \phi^4 \right).
\end{equation}
The corresponding translation Ward identity (TWI) for a generic probe $P$ reads
\begin{equation}
 \langle\, \delta_{x,\rho}P \,\rangle = -\langle\, P\;\partial_{\mu}T_{\mu\rho}(x) \,\rangle,
 \label{eq:TWI}
\end{equation}
where the local operator of translation is defined as \cite{DelDebbio:2013zaa}
\begin{equation}
 \delta_{x,\rho}P = \frac{\delta P}{\delta\phi(x)}\,\partial_{\rho}\phi(x).
\end{equation}
The TWI implies that $\partial_{\mu}T_{\mu\rho}$ is finite as the left-hand side of eq. (\ref{eq:TWI}) is finite. Hence, any divergence in the EMT has to be transverse, i.e. proportional to $(\partial_\mu\partial_\rho-\delta_{\mu\rho}\partial^2)\phi^2$. However, inserting a term of this form into $n$-point functions cannot produce any divergence. Firstly, the derivatives lead only to external momenta. Secondly, $\phi^4$-theory in three dimensions is superrenormalisable. Studying the superficial degree of divergence shows that there exist only two divergent diagrams, one is linear divergent, one is logarithmic divergent. The insertion of $\phi^2$ is finite as it only leads to an additional propagator in loop diagrams and hence renders both divergent diagrams finite. Thus, $T_{\mu\rho}$ itself is finite.

\section{Lattice energy-momentum tensor}

The action on the lattice is
\begin{equation}
 \hat{S} = a^3\sum_n \left( \frac12\hat{\partial}^F_{\mu}\phi\hat{\partial}^B_{\mu}\phi + \frac12m_0^2\phi^2 + \frac{\lambda_0}{4!}\phi^4 \right),
\end{equation}
where the hat indicates lattice quantities, and $\hat{\partial}^F_{\mu}\phi$ and $\hat{\partial}^B_{\mu}\phi$ are forward and backward derivatives, respectively. Lattice regularisation breaks translation symmetry explicitly. Thus, there is an extra term $\hat{R}_{\rho}$ appearing in the TWI,
\begin{equation}
 \langle\, \hat{\delta}_{x,\rho} \hat{P} \,\rangle = - \langle\, \hat{P} \left(\hat{\partial}^S_{\mu}\hat{T}_{\mu\rho} + \hat{R}_{\rho} \right)\,\rangle,
\end{equation}
where $\hat{T}_{\mu\rho}$ is the naively discretised EMT, and $\hat{\partial}^S_{\mu}$ is the symmetric derivative. It is important to use $\hat{\partial}^S_{\mu}$ as it conserves the unitarity of the transformation $\phi(x)\rightarrow\phi(x')=\phi(x)+\alpha\hat{\partial}^S_{\mu}\phi(x)=\exp(\alpha\hat{\partial}^S_{\mu})\phi(x)$. $\hat{R}_{\rho}$ is made of irrelevant operators that are suppressed by powers of the lattice cutoff. However, these powers can combine with divergences $1/a$ appearing in expectation values to give finite contributions. Thus, the EMT has to be renormalised. The renormalised lattice TWI reads
\begin{equation}
 \langle\, Z_{\delta} \;\hat{\delta}_{x,\rho} \hat{P} \,\rangle = - \langle\, \hat{P} \left(\hat{\partial}^S_{\mu}[\hat{T}_{\mu\rho}] + \hat{\bar{R}}_{\rho} \right)\,\rangle,
\end{equation}
where $\hat{\bar{R}}_{\rho}$ is finite. The renormalised $\hat{T}_{\mu\rho}$ is the sum of all possible terms that can mix with the EMT, and their expectation values,
\begin{equation}
 [\hat{T}_{\mu\rho}(x)] = \sum_i c_i \left\{ \hat{T}_{\mu\rho}^{(i)} - \langle \hat{T}_{\mu\rho}^{(i)} \rangle \right\}.
 \label{eq:renormEMTgeneral}
\end{equation}
Possible mixing terms need to be of mass dimension equal or smaller than three, Lorentz-invariant, and invariant under $\phi\rightarrow-\phi$ and $x\rightarrow-x$. We find ten mixing terms,
\begin{equation}
 \hat{\partial}_{\mu}\phi\hat{\partial}_{\rho}\phi,\; \; \phi\hat{\partial}_{\mu}\hat{\partial}_{\rho}\phi,\;\; \delta_{\mu\rho}\left( 1,\;\; \phi^2,\;\;  \phi^4,\;\;  \phi^6,\;\;  \sum_{\sigma}\hat{\partial}_{\sigma}\phi\hat{\partial}_{\sigma}\phi,\;\;  \sum_{\sigma}\phi\hat{\partial}_{\sigma}\hat{\partial}_{\sigma}\phi,\; \; \hat{\partial}_{\mu}\phi\hat{\partial}_{\mu}\phi,\; \; \phi\hat{\partial}_{\mu}\hat{\partial}_{\mu}\phi \right).
\end{equation}
A perturbative analysis shows that any occurring divergence is proportional to $\phi^2$. Hence, there are no new terms appearing in the renormalised EMT and the only coefficient that needs to change in order to render the EMT finite is the one corresponding to the $\phi^2$-term. Thus, the renormalised EMT reads
\begin{equation}
 [\hat{T}_{\mu\rho}] = \frac{c}{2}  \hat{\partial}_{\mu}\phi \hat{\partial}_{\rho}\phi + \delta_{\mu\rho}\left( \frac{c_2}{2} \phi^2 + \frac{c'}{2} \sum_{\sigma} \hat{\partial}_{\sigma}\phi \hat{\partial}_{\sigma}\phi + \frac{c_4}{4!} \phi^4 \right),
 \label{eq:renormEMT}
\end{equation}
and we expect for the coefficients $c=2$, $c_2/m_0^2=0.83$, $c'=-1$, $c_4/\lambda_0=-1$. The value of $c_2$ consists of the sum of the bare mass and the term that cancels the divergence caused by the bare TWI. We determined it using perturbation theory on the lattice.

In principle, one can now compute the renormalised EMT on the lattice, but contact terms complicate the task. A more numerically feasible method is the use of the Wilson flow.

\section{The Wilson flow - gradient flow on the lattice}

\begin{figure}[h]
\centering
 \includegraphics[width=4.5cm]{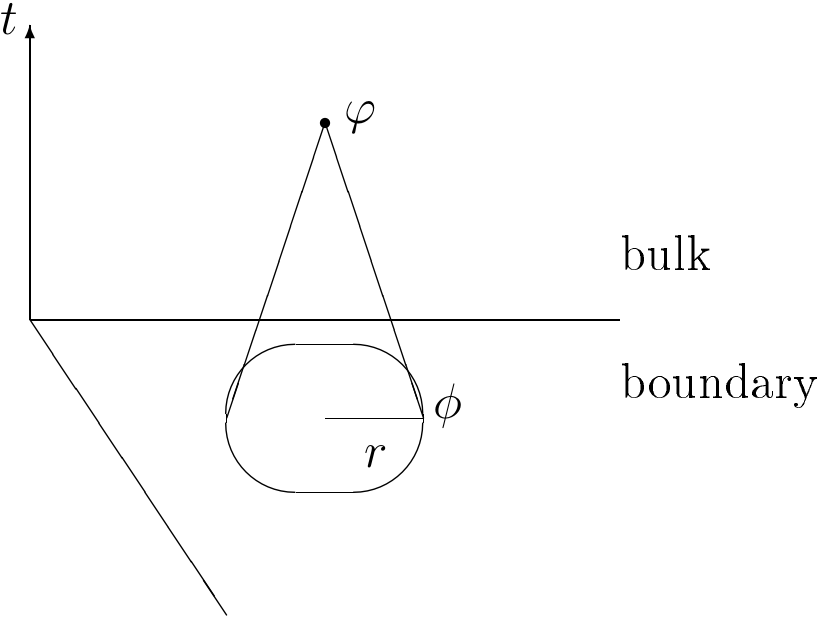}
 \label{fig:Wflow}
\end{figure}

The flow equation for scalar field theory was first formulated in \cite{Monahan:2015lha},
\begin{equation}
 \partial_t\varphi(t,x) = \partial^2 \varphi(t,x), \qquad \left.\varphi(t,x)\right\rvert_{t=0} = \phi(x).
 \label{eq:Wflow}
\end{equation}
The flow time $t$ and a new field, the flow field $\varphi$, were introduced. The flow equation determines the evolution of $\varphi$ along the flow. The flow has a smearing effect on the boundary fields $\phi$ with smearing radius $r=\sqrt{6t}$. Fig. (\ref{fig:Wflow}) illustrates the situation: the scalar field lives at the boundary at $t=0$, the flow field lives in the bulk. The bulk extents from the boundary along direction $t$.

\section{Renormalisation of the EMT using the Wilson flow}
\label{sec:renormEMTflow}

In order to renormalise the EMT on the lattice we use probes at non-zero flow time $T$ instead of probes at the boundary. Doing so avoids the appearance of contact terms. The renormalised TWI then reads
\begin{equation}
 \langle\, Z_{\delta} \;\hat{\delta}_{x,\rho} \hat{P_T} \,\rangle = - \langle\, \hat{P}_T \left(\hat{\partial}^S_{\mu}[\hat{T}_{\mu\rho}] + \hat{\bar{R}}_{\rho} \right)\,\rangle.
\end{equation}
The coefficients in eq. (\ref{eq:renormEMTgeneral}) can be tuned such that the EMT is finite and that $\hat{\bar{R}}_{\rho}$ vanishes in the continuum limit,
\begin{equation}
 Z_{\delta}\, \langle\, \hat{\delta}_{x,\rho} \hat{P}_T \,\rangle = - \sum_i c_i \;\langle\, \hat{P}_T \;\hat{\partial}^S_{\mu}\hat{T}_{\mu\rho}^{(i)}(x) \,\rangle
\end{equation}
$Z_{\delta}$ can be determined separately and we find $Z_{\delta}=1$.

The remaining task is to solve a system of equations with different operators $P^{(j)}_T$,
\begin{equation}
 \langle\, \hat{\delta}_{x,\rho} \hat{P}_T^{(j)} \,\rangle = - \sum_i c_i \;\langle\, \hat{P}_T^{(j)} \;\hat{\partial}^S_{\mu}\hat{T}_{\mu\rho}^{(i)}(x) \,\rangle,
 \label{eq:system}
\end{equation}
where the index $j$ counts the number of different probes.

\section{Results}

We used Monte Carlo simulations to study the scalar EMT non-perturbatively. In particular, we used the algorithm in \cite{Brower:1989mt} which consists of local Metropolis updates alternating with Swendsen Wang cluster updates of the embedded Ising model. Autocorrelation was monitored and found negligible for every simulation. We implemented the Wilson flow, eq. (\ref{eq:Wflow}), using an integration routine based on the fourth order Runge-Kutta method. In principle, in scalar theory it is possible to integrate the flow equation exactly but implementing the result is very expensive.

\begin{figure}
\centering
 \includegraphics[width=8.5cm,trim={1cm 1cm 1cm 3cm},clip]{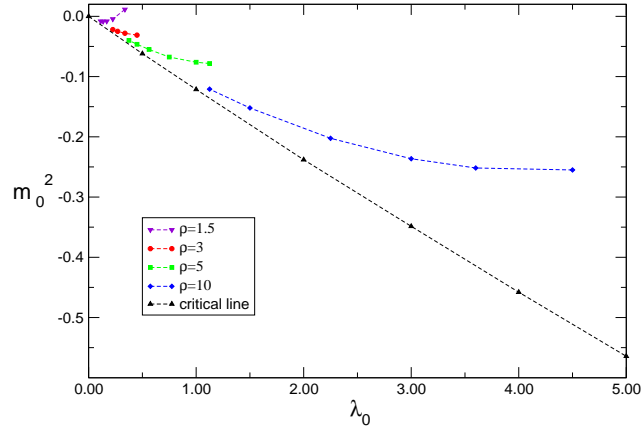}
 \caption{Critical line and lines of constant physics in scalar $\phi^4$-theory.}
\label{fig:rentraj}
\end{figure}

We are interested in staying close to critical line in the symmetric phase. The line of second order phase transition separates the broken from the symmetric phase. We determined it by examining the peak of the susceptibility, fixing $\lambda_0$ and varying $m_0$. The critical line for small values of $\lambda$ is shown in fig. \ref{fig:rentraj}. In order to study the continuum limit of the EMT we have to be close to the ultraviolet fixed point and follow lines of constant physics. We defined lines of constant physics via the ratio
\begin{equation}
 \rho = \frac{\lambda_0}{m_R}.
\end{equation}
We can use the bare coupling $\lambda_0$ instead of the renormalised one because the theory is superrenormalisable and so $\lambda_0=\lambda_R+\mathcal{O}(a)$. The lines of constant physics we found for $\rho=1.5, 3, 5, 10$ are drawn in fig. \ref{fig:rentraj}. We set the lattice spacing by choosing a value for $a\lambda_0$. Then, we search for $am_0$ such that the value of $\rho$ is constant. We kept the physical volume constant and increased the fineness of the lattice so that the continuum is approached by following each line toward the critical line and ultraviolet fixed point. An observable can then be evaluated along these lines and extrapolated to the continuum.

\begin{figure}
\centering
 \includegraphics[width=0.47\textwidth,trim={0cm 1.5cm 2cm 3cm},clip]{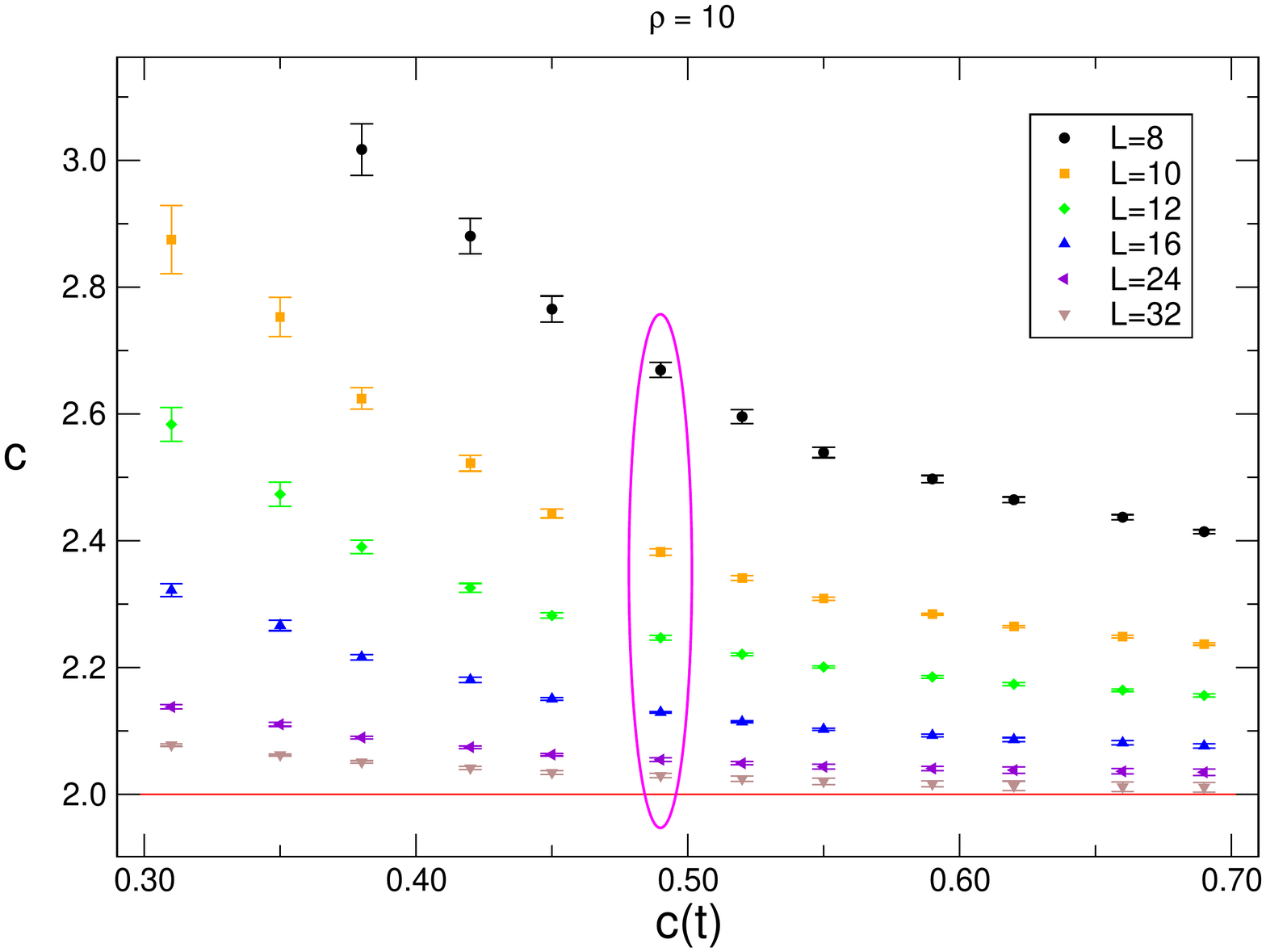}\hspace{0.5cm}
 \includegraphics[width=0.47\textwidth,trim={0cm 1.5cm 2cm 3cm},clip]{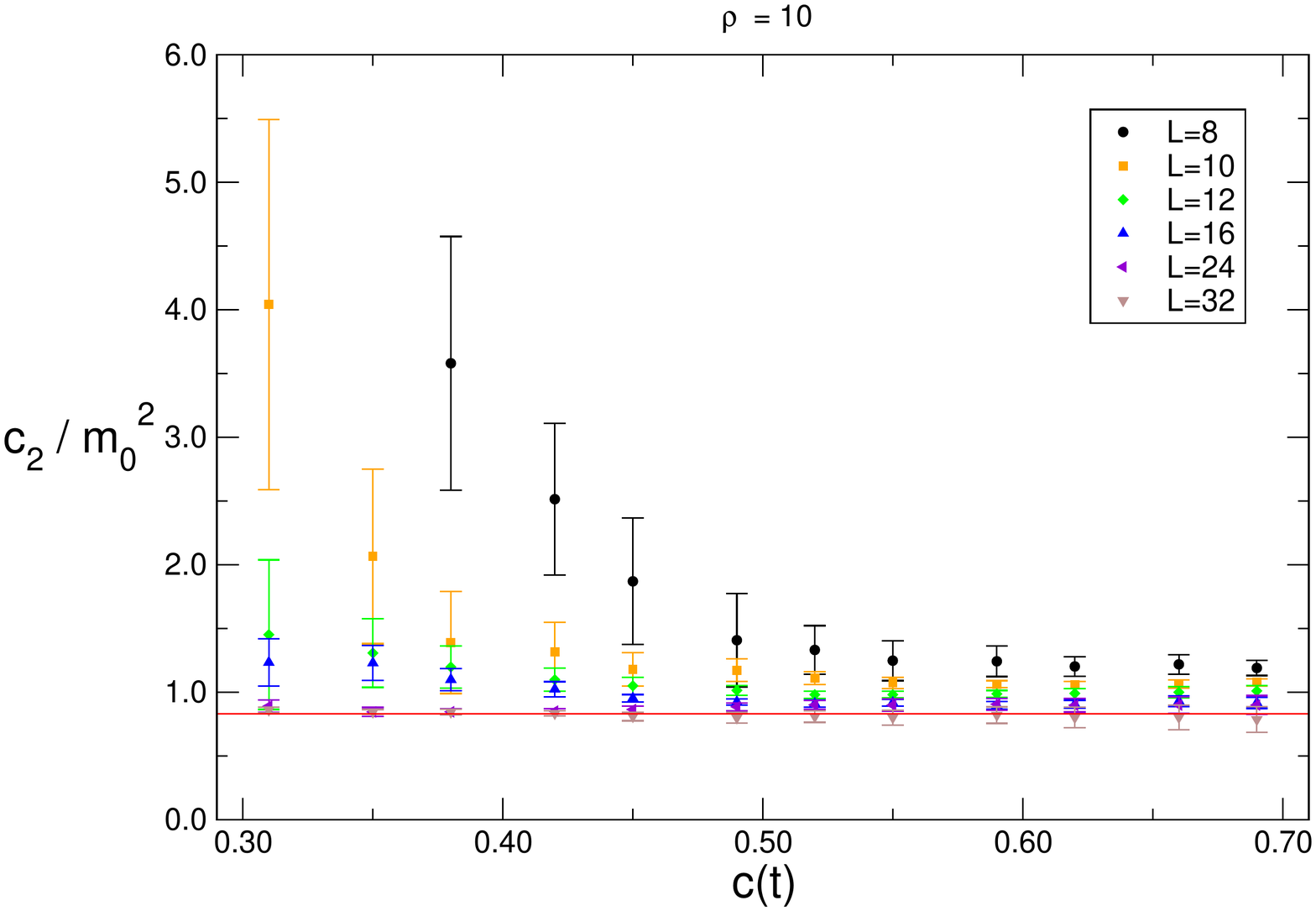}\vspace{0.5cm}
 \includegraphics[width=0.47\textwidth,trim={0cm 1.5cm 2cm 3cm},clip]{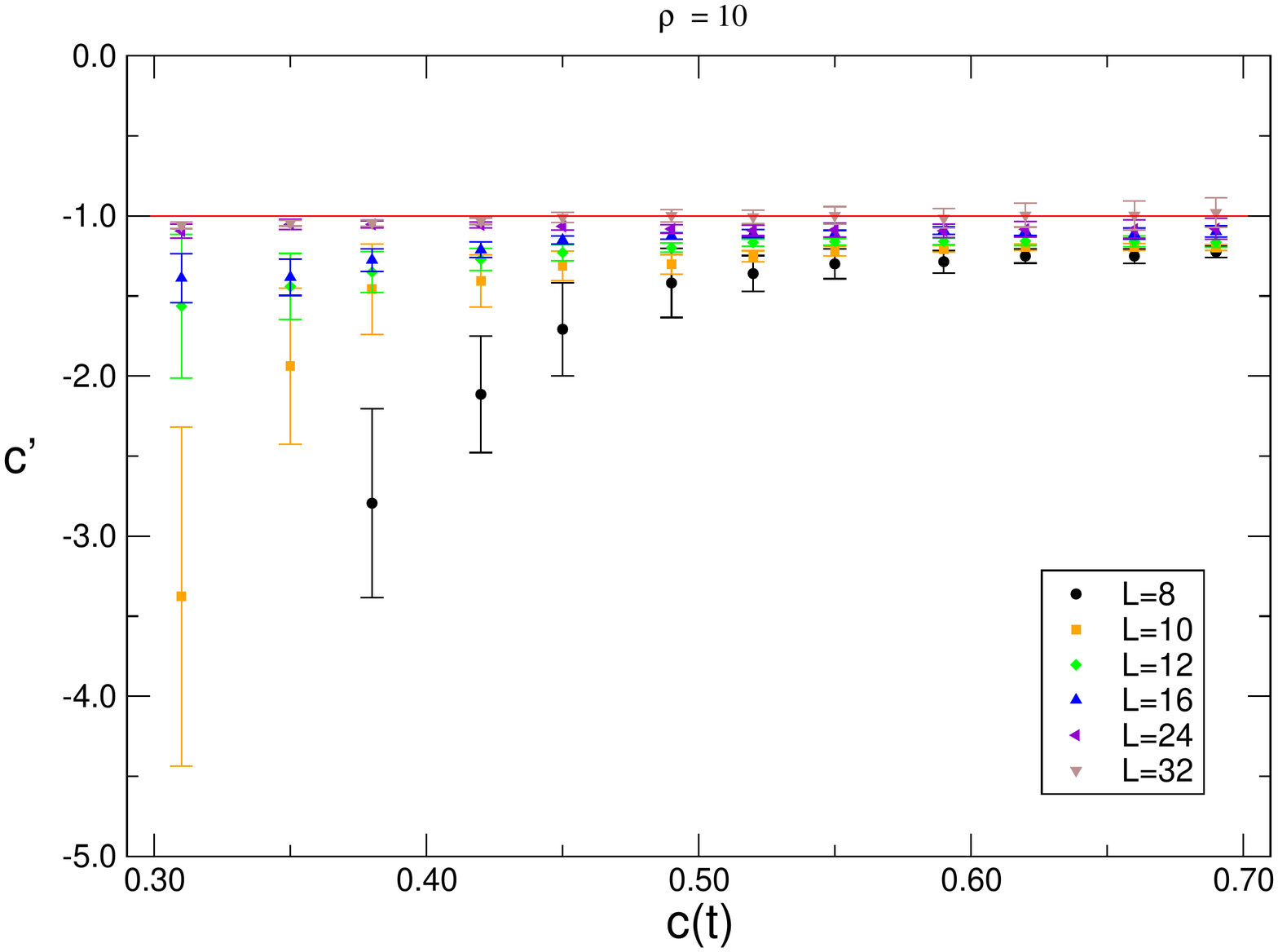}\hspace{0.5cm}
 \includegraphics[width=0.47\textwidth,trim={0cm 1.5cm 2cm 3cm},clip]{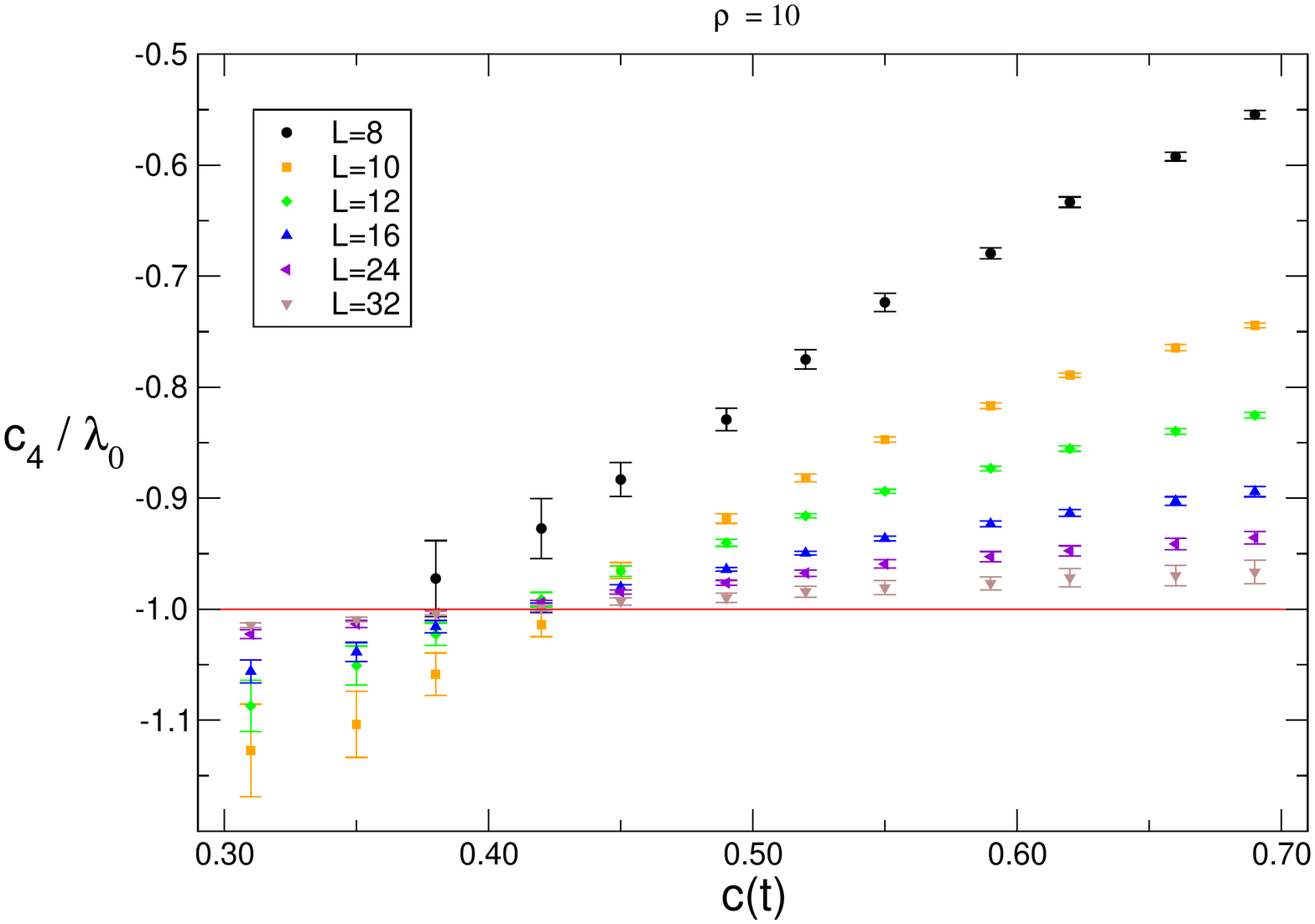}
 \caption{Results for the coefficients of the renormalised EMT for $\rho=10$.}
\label{fig:c1234}
\end{figure}

In order to determine the coefficients in eq. (\ref{eq:renormEMT}) we need to invert the system, eq. (\ref{eq:system}). It turns out that using an overdetermined system is a better choice than using a square system in terms of efficiency and range of validity. As discussed in sec. \ref{sec:renormEMTflow} we already know what to expect for the coefficients. The results shown in fig. \ref{fig:c1234} for $c$, $c_2/m_0^2$, $c'$ and $c_4/\lambda_0$ for $\rho=10$ for different lattice spacings and flow times $c(t)=\sqrt{6t}/L$ confirm our expectations. At small flow times the measurements are dominated by lattice artefacts. Where $c(t)\rightarrow1$ the smearing radius equals the lattice size. Thus, we expect reasonable results in between these two flow times. With finer lattices a plateau is forming and we can already see that the expected values are approached towards the continuum, for all flow times.

\begin{figure}
\centering
 \includegraphics[width=0.48\textwidth,trim={1.5cm 1.2cm 1.5cm 3cm},clip]{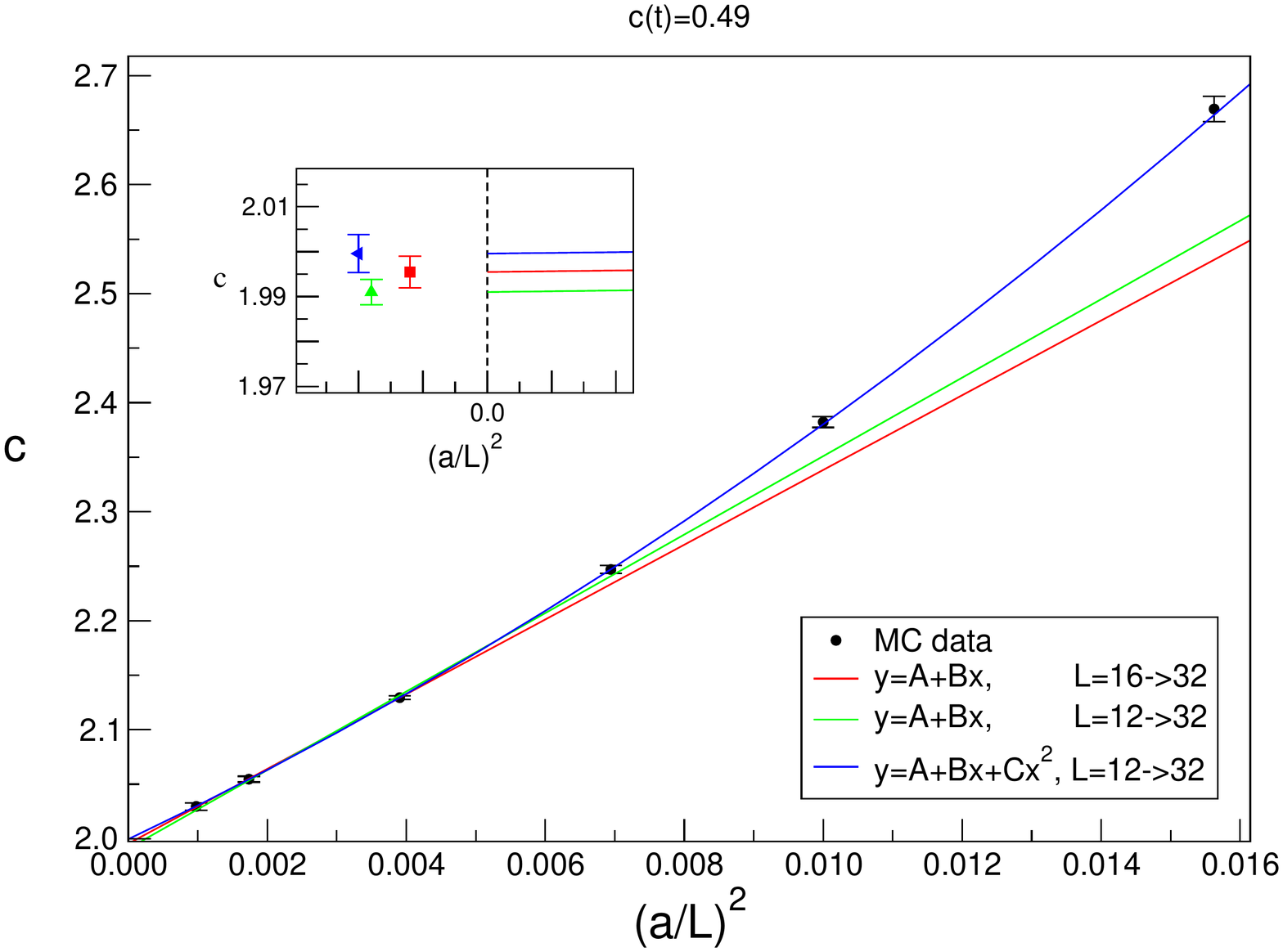}\hspace{0.5cm}
 \includegraphics[width=0.48\textwidth,trim={1.5cm 1.2cm 1.5cm 3cm},clip]{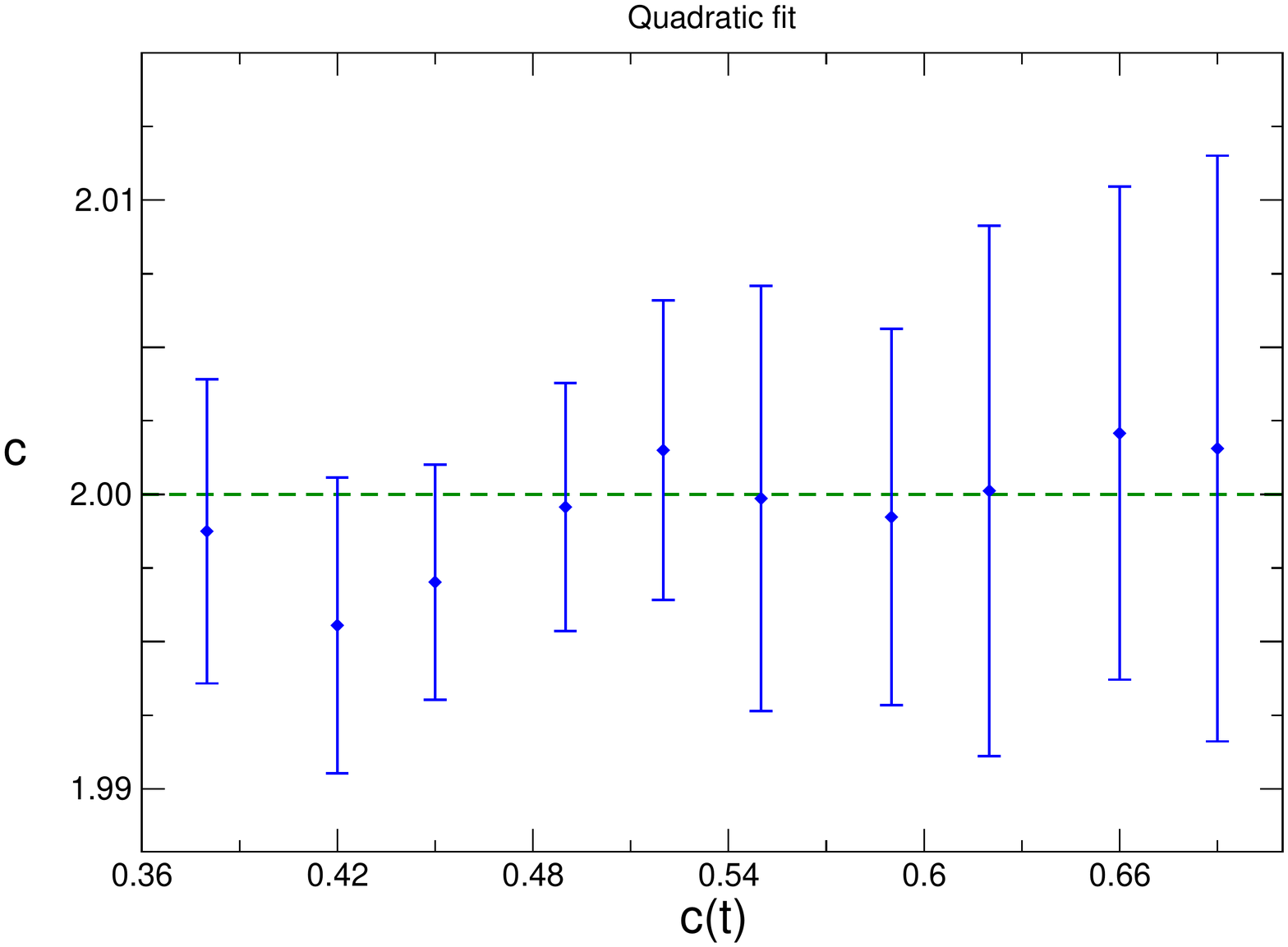}
 \caption{Left: Continuum extrapolation of $c$ for $\rho=10$ and $c(t)=0.49$. Right: Continuum values of $c$ at different flow times using the quadratic fit in the left diagram.}
\label{fig:continuumfit}
\end{figure}

As an example, a proper fit for the continuum value of $c$ at $c(t)=0.49$ encircled in pink in fig. \ref{fig:c1234} is displayed in fig. \ref{fig:continuumfit}. In the left diagram $c$ is plotted as a function of $(a/L)^2$. We made three different fits, two linear ones, one including only the last three points, one including only the last four points, and a quadratic fit including the last four points. The reduced $\chi^2$ is 0.054, 3.04 and 0.024, respectively, revealing that the quadratic fit connects the data best. The small plot within shows the region close to the continuum. The second plot on the right shows the extrapolation of $c$ at different flow times to the continuum using the quadratic fit.

\section{Conclusion}

We studied the non-perturbative renormalisation of the EMT using the Wilson flow. The Wilson flow provides a new way to implement Ward identities free from contact terms. With its help we are able to define a properly renormalised EMT on the lattice and find the coefficients of the renormalised EMT at finite lattice spacing. Each flow time gives a different definition of the EMT. We are also able to reproduce the correct continuum limit of the EMT. With this new knowledge we now have the means to study the trace of the EMT and thus the scaling behaviour of the scalar theory.

\section*{Acknowledgements}

We thank A. Patella, L. Keegan, K. Skenderis and M. Hanada for fruitful discussions. The numerical computations have been carried out using resources from the HPCC Plymouth. LDD is supported by STFC, grant ST/L000458/1, and the Royal Society, Wolfson Research Merit Award, grant WM140078. SE is supported by the  Higgs Centre for Theoretical Physics. RP and AP are supported by STFC (grant ST/L000458/1). AR is supported by the Leverhulme Trust (grant RPG-2014-118) and STFC (grant ST/L000350/1).

\end{document}